\documentclass[aps, prd, twocolumn, superscriptaddress, nofootinbib, showpacs,
floatfix]{revtex4}
\usepackage[dvips]{epsfig}

\def\lsim{\lower0.6ex\vbox{\hbox{$ \buildrel{\textstyle <}\over{\sim}\ $}}}
\def\gsim{\lower0.6ex\vbox{\hbox{$ \buildrel{\textstyle >}\over{\sim}\ $}}}

\def\beq{\begin{equation}}
\def\eeq{\end{equation}}

\def\beqa{\begin{eqnarray}}
\def\eeqa{\end{eqnarray}}

\def\bfig{\begin{figure}[ht] \begin{center}}
\def\bfigh{\begin{figure}[h!] \begin{center}}
\def\bfigb{\begin{figure}[hb!] \begin{center}}
\def\bfigt{\begin{figure}[t!] \begin{center}}
\def\bfight{\begin{figure}[ht!] \begin{center}}
\def\efig{\end{center} \end{figure}}

\def\btab{\begin{table*}[ht]}
\def\etab{\end{table*}}

\def\mnras{{Mon. Not. R. Astron. Soc.}~}

\def\etal{{\it et al.}~}

\def\ie{{\it i.e.},~}

\begin{document}

\title{The Evolution of Inverse Power Law Quintessence at Low Redshift}

\author{Casey R. Watson}
\affiliation{Department of Physics, The Ohio State University,
Columbus, OH 43210, USA}
\email{cwatson@pacific.mps.ohio-state.edu}
\author{Robert J. Scherrer}
\affiliation{Department of Physics, The Ohio State University, 
Columbus, OH 43210, USA}
\affiliation{Department of Astronomy, The Ohio State University,
Columbus, OH 43210, USA}
\email{scherrer@pacific.mps.ohio-state.edu}

\date{\today}

\begin{abstract}

Quintessence models based on a scalar field, $\phi$, with an
inverse power law potential display simple tracking behavior
at early times, when the quintessence energy density, $\rho_\phi$, 
is sub-dominant.
At late times, when $\rho_\phi$ becomes
comparable to the matter density, $\rho_m$,
the evolution of $\phi$ diverges from its scaling behavior.
We calculate the first order departure of $\phi$ from 
its tracker solution at low redshift.
Our results for the evolution of $\phi$,
$\rho_\phi$, $\Omega_\phi$, and $w$ are
surprisingly accurate even down to $z=0$.  We find
that $w$ and $\Omega_\phi$ are related linearly 
to first order, and derive a semi-analytic expression for $w(z)$ which
is accurate to within a few percent.  Our analytic techniques are potentially applicable
to any quintessence model in which the quintessence component comes to dominate at late
times.

\end{abstract}

\pacs{98.80.Cq}

\maketitle

\section{\label{sec:Intro}Introduction}

A host of observations suggest that the universe is currently dominated by some
form of energy with
negative pressure \cite {flat,WMAP,lowM,SN}.
Recent
observations of Type Ia supernovae
\cite{SN}, which appear fainter than they would if the expansion were
decelerating due to matter alone, provide the
most direct evidence for a dominant negative pressure component (and the
accelerating expansion that accompanies
it).  In addition, cosmic microwave background observations from
WMAP, for example, suggest
that the universe is nearly flat ($\Omega_0 \approx 1$) but contains a
sub-critical density of matter $\Omega_{m0} = 0.27 \pm 0.04$ \cite{WMAP}.  Taken
together these
findings imply the presence of a large fraction of unclustered ``dark energy". 

The most straightforward source of dark energy is a cosmological constant, but the
level of fine-tuning necessary to make it a viable candidate is disconcerting. 
An alternative, dubbed quintessence, is a model of time-varying dark energy based on 
the behavior of a dynamical,
classical scalar
field $\phi$.  In the simplest quintessence models, $\phi$ is taken to be a
minimally-coupled field
with potential $V(\phi)$, and
pressure, energy density, and equation of state
\beq
   p_\phi = {1\over 2} \dot\phi^2 - V(\phi),
\label{eq:pphi}
\eeq
\beq
   \rho_\phi = {1\over 2} \dot\phi^2 + V(\phi),
\label{eq:rhophi}
\eeq
and
\beq
   w = p_\phi/\rho_\phi.
\label{eq:wphi}
\eeq

The behavior of the quintessence field will depend, of course, on the particular
choice of $V(\phi)$.
One widely-investigated case is the inverse power law potential \cite{IPL}
\beq
V(\phi) = \kappa\phi^{-\alpha},
\label{eq:Vphi}
\eeq
where $\kappa$ is a constant with units of $m^{4+\alpha}$.

These potentials are particularly simple and have several desirable properties.
They exhibit tracking behavior, $\ie$ a wide
range of initial conditions converge to a common solution at late times
\cite{Tracker,lowIPL}. In addition, inverse power law
quintessence remains sufficiently subdominant energetically so that it does not
disastrously interfere with known
epochs of standard big bang cosmology. Finally, because a fundamental
understanding of the nature of
dark energy must eventually come from particle physics, it is worth noting that
inverse power law potentials can arise naturally
from extensions to the standard model (e.g., SUSY QCD \cite{SUSYQCD}).

The inverse power law tracker solutions are characterized by an equation of state parameter,
$w$, which is approximately
constant when the universe is dominated by a background fluid.  During the
matter-dominated era,
for example, when the contribution of the quintessence energy
density to the expansion is neglected,
$w$ is given by \cite{IPL,Tracker,lowIPL}:
\begin{equation}
w = - {2 \over 2+\alpha},
\end{equation}
and $\phi$ evolves as
\begin{equation}
\phi \propto t^{2/(2+\alpha)},
\end{equation}
corresponding to a density which evolves as
\begin{equation}
\rho_\phi \propto t^{-2 \alpha/(2+\alpha)}.
\end{equation}
These expressions provide an excellent approximation to the behavior
of the quintessence field as long as $\rho_\phi/\rho_m \ll 1$.

At late times, however, when the scalar field itself contributes
significantly to the
expansion of the universe,
the value of $w$ begins to diverge from its tracker value, as
do $\phi(t)$ and $\rho_\phi(t)$.   In this paper, we
examine analytically
the behavior of the scalar field during the transition from matter to
scalar-field domination.
It is this late-time evolution which is relevant to supernova Ia observations
and other tests of dark energy
(see, e.g., Ref. \cite{multiObs} and references therein).  A good analytic
approximation to the behavior
of $\Omega_\phi$ and $w$ at late times would be useful in such calculations. 
Furthermore, our treatment
provides insight into the behavior of the scalar field once it begins to
dominate.

A number of observations provide stringent upper limits on $w$. 
For example, the WMAP results, combined with supernovae observations, HST data,
2dFGRS observations of large scale 
structure, and Lyman alpha forest data give
$w \lesssim -0.78$ at 2 $\sigma$ \cite{WMAP}, assuming constant
$w$, which implies a low value
for $\alpha$.  Nonetheless, we have attempted to keep our approach as general
as possible, treating
all values of $\alpha$ in the discussion below.

In the following section, we rederive the inverse
power law model solutions for
$\phi$, $\rho_{\phi}$,
and $w$ during the matter-dominated era.  In Section~\ref{sec:PMDE}, we
use the
first-order perturbations to these background solutions to describe the deviation
of the scalar field
away from tracking behavior
as the universe enters the scalar-field dominated era.
In Section~\ref{sec:NUM}, we compare our analytical results to those of the
numerically-integrated evolution equations.  We find that the first-order
corrections to the background solutions
for $\phi$, $\rho_{\phi}$, $\Omega_{\phi}$, and $w$ characterize their
behavior surprisingly well even to the present.  
In Section~\ref{sec:CON}, we discuss some consequences
of these results.
	
\section{\label{sec:MDE} Inverse Power Law Quintessence Evolution in the
matter-dominated era}

The equation of motion for $\phi$ is
\beq
\ddot\phi + 3H\dot\phi + {dV(\phi)\over d\phi } = 0,
\label{eq:eom}
\eeq
where $H = \dot a/ a$ is the Hubble parameter and $a = a(t)$ is the scale
factor which describes the
expansion of the universe as a function of time.

We confine our analysis to
$z < 30$, so that the contribution of radiation to the expansion
can be neglected.
Assuming a flat universe containing only matter and quintessence,
the Friedmann equation is 
\beq
3H^2 = \rho_{m} + \rho_{\phi},
\label{eq:H}
\eeq
in units where $8\pi G = 1$, which we will use throughout the
paper.

When $\rho_\phi$ is neglected in equations (\ref{eq:eom}) and (\ref{eq:H}),
it is easy to integrate these equations directly to derive $\phi(t)$,
as was done in Refs. \cite{IPL,Tracker,lowIPL}.  However, in deriving
a perturbative expansion when $\rho_\phi/\rho_m$ is small but nonzero, it turns
out to be easier to use $a$ as the independent variable, rather than $t$.
Further, we would like to express observable
quantities such as $w$ and $\Omega_\phi$
at late times as functions of redshift, rather than time, which is
straightforward when the independent variable is taken to be $a$.

Appealing to equation (\ref{eq:rhophi}) and equation~(\ref{eq:H}), 
${1\over 2} \dot\phi^2$ can be rewritten as
\beq
{1\over 2} \dot\phi^2 = {x\over 1-x}(\rho_m + V),
\label{eq:phidota} 
\eeq 
where $V$ implicitly means $V(\phi)$ and 
\beq x = {\rho_\phi + p_\phi\over 2(\rho_m + \rho_\phi)} = {{1\over 2} \dot\phi^{2}\over 3H^2} = 
{1\over 6}{\Big(a{d\phi\over da}\Big)}^2.
\label{eq:x} 
\eeq 
Substituting equation~(\ref{eq:phidota}) into Eqs. (\ref{eq:pphi}), (\ref{eq:rhophi}), (\ref{eq:wphi}), and 
(\ref{eq:H}), we can rewrite $\rho_{\phi}$, $w$, and the Friedmann equation as follows:
\beq
\rho_{\phi} = {{x\rho_m + V}\over 1-x},
\label{eq:rhophia} 
\eeq 
\beq
w = {{x\rho_m - V(1-2x)}\over {x\rho_m + V}},
\label{eq:wphia} 
\eeq 
and
\beq
3H^2 = {{\rho_{m} + V}\over 1-x}.
\label{eq:Ha} 
\eeq 

Using the above expressions, equation~(\ref{eq:eom}) becomes
\beq 
a^{2}{\phi}'' + {a{\phi}'\over 2}(5-3x) + {3(1-x)\over \rho_m +V}\Big({a{\phi}'V\over 2} 
+ {dV\over d\phi}\Big) = 0,
\label{eq:eoma} 
\eeq 
where the prime denotes $d/da$.
During the matter-dominated epoch, when $x \ll 1$, equation (\ref{eq:eoma}) 
reduces to 
\beq 
a^{2}{\phi}''_{(0)} + {5a{\phi}'_{(0)}\over 2} + {3\over \rho_m}{dV\over d\phi} = 0, 
\label{eq:eomamde} 
\eeq 
where the zero subscript in parentheses
denotes the zeroth-order solution, neglecting
the contribution of quintessence to the expansion rate.  (We
use the subscript $``(0)"$ throughout to denote such zeroth-order terms,
while the subscript $``0"$ without parentheses is reserved for quantities
evaluated at the present day).

For quintessence with an inverse power law potential (equation~\ref{eq:Vphi}),
the solution to equation~(\ref{eq:eomamde}) is
\beq
\phi_{(0)} = C(\alpha){a}^{{3/(2 + \alpha)}}.
\label{eq:bckgrndphia}
\eeq

In equation (\ref{eq:bckgrndphia}), the constant C($\alpha$) depends on the
value of $\alpha$
and
is related to $\kappa$ and the present matter density, $\rho_{m0}$, in the following way
\beq
\kappa = \rho_{m0}{3(4+\alpha)\over 2\alpha (2+\alpha)^{2}}{C(\alpha)}^{2+\alpha}.
\label{eq:constC}
\eeq
The corresponding zeroth-order solution for $\rho_\phi$ 
is
\beq
\label{phidensity}
\rho_{\phi (0)} = 3\rho_{m0}{{C(\alpha)^2}\over {\alpha(2 + \alpha)}}{a}^{-{3\alpha/(2 + \alpha)}}.
\label{eq:bckgrndrhophia}
\eeq
The zeroth-order behavior of the quintessence density parameter is then given by 
$\Omega_{\phi (0)} = \rho_{\phi (0)}/(\rho_{\phi (0)} + \rho_{m})$.
Finally, the equation of state is just a constant in the tracking regime:
\beq
\label{w(0)}
   w_{(0)} = -{2\over 2+\alpha}. 
\label{eq:bckgrndwphia} 
\eeq
These expressions for $\phi_{(0)}(a)$, $w_{(0)}(a)$, and $\rho_{\phi (0)}(a)$
are, of course, equivalent to the previously-derived time-dependent quantities
given in Section I.

\section{\label{sec:PMDE}First-Order Perturbation to Inverse
Power Law Quintessence Evolution}

We define the first-order perturbation to the zeroth-order background field $\phi_{(0)}$ 
(equation~\ref{eq:bckgrndphia}) to be
$\phi_{(1)}$, where
\beq
\phi_{(1)}/\phi_{(0)} \sim O(\rho_{\phi (0)}/\rho_m).
\eeq
Then the perturbed field, $\phi$, is, to first order, 
\beq 
\widetilde\phi = \phi_{(0)} + \phi_{(1)},
\label{eq:fieldpert} 
\eeq 
where we use a tilde throughout to denote quantities expanded to first
order in $\rho_{\phi(0)}/\rho_m$.
Substituting equation~(\ref{eq:fieldpert}) into equation~(\ref{eq:eoma}) and keeping all terms of order
${\rho_{\phi(0)}/ \rho_m}$ (note, in particular, that $x\sim O(\rho_{\phi (0)}/\rho_m)$), we find that $\phi_{(1)}$ obeys 
\beqa 
a^{2}{\phi_{(1)}''} + {a\over 2}(5\phi_{(1)}'-3x_{(0)}\phi_{(0)}') + {3a\phi_{(0)}'V\over 2 \rho_{m}}
\nonumber \\
- {3\rho_{\phi (0)}\over  {(\rho_{m})}^{2}} {dV\over d\phi}+ {3 \over \rho_{m}}{d^2 V \over
d\phi^2}\phi_{(1)} = 0,
\label{eq:perteomamdegen} 
\eeqa 
where $x_{(0)} = (a\phi'_{(0)})^{2}/6$.

For our particular choice of potential, equation~(\ref{eq:perteomamdegen}) becomes
\beqa
a^{2}{\phi_{(1)}''} &+& {a\over 2}(5\phi_{(1)}'-3x_{(0)}\phi_{(0)}') + 
{27 \kappa a^{3}\over 2(2+\alpha)\rho_{m0}{(\phi_{(0)})^{(-1+\alpha)}}} 
\nonumber \\
&+& {3\alpha(1+\alpha)\kappa a^{3} \phi_{(1)}\over \rho_{m0}{(\phi_{(0)})^{(2+\alpha)}}} = 0.
\label{eq:perteomamde}
\eeqa
The solution to  equation~(\ref{eq:perteomamde}) is
\beq
\phi_{(1)} =  -{3(6+\alpha)\phi_{(0)}^{3}\over \alpha(2+\alpha)(28+8\alpha + {\alpha}^{2})}.
\label{eq:phi1soln}
\eeq

The corresponding corrections to $\rho_{\phi (0)}$ and $w_{(0)}$ are
\beq
\rho_{\phi (1)} = x_{(0)}\rho_{\phi (0)} + x_{(1)}\rho_m - \alpha{\phi_{(1)}\over \phi_{(0)}}V,
\label{eq:rhopert}
\eeq
and
\beq
w_{(1)} = (1-w_{(0)})\Big({\rho_{\phi (1)}\over \rho_{\phi (0)}} + 
\alpha{\phi_{(1)}\over \phi_{(0)}}\Big),
\label{eq:wpert}
\eeq
where $x_{(1)} = 2x_{(0)}{\phi_{(1)}'/\phi_{(0)}'}$
is defined via
$\widetilde{x} = x_{(0)} + x_{(1)} = a^{2}(\phi'^{2}_{(0)} + 2\phi'_{(0)}\phi'_{(1)})/6$.

In terms of $\phi_{(0)}$ and $\phi_{(1)}$, the first-order 
perturbations to $\rho_{\phi (0)}$ and $w_{(0)}$ are
\beq
\rho_{\phi (1)} = -\frac{\alpha(4+\alpha)} {6+\alpha}\Big(
{\phi_{(1)}\over \phi_{(0)}}\Big)\rho_{\phi (0)},
\label{eq:rhopertsolna}
\eeq
and 
\beq
w_{(1)} = -\frac{\alpha(4+\alpha)} {6+\alpha}\Big({\phi_{(1)}\over \phi_{(0)}}\Big)w_{(0)}.
\label{eq:wpertsolna}
\eeq
Because $\phi_{(1)}$ is negative, equation~(\ref{eq:rhopertsolna}) and
equation~(\ref{eq:wpertsolna}) predict that $\rho_{\phi (1)}$ is positive
and $w_{(1)}$ is negative, as expected.  Note also that
$w_{(1)}/w_{(0)} = \rho_{\phi(1)}/\rho_{\phi(0)} = p_{\phi(1)}/2p_{\phi(0)}$.

Combining the zeroth-order and first-order solutions for
$\phi$, $\rho_{\phi}$, and $w$, we get
\begin{eqnarray}
\widetilde \phi &=&
\phi_{(0)} +  \phi_{(1)},\nonumber\\
 &=& \phi_{(0)} - {3 (6 + \alpha) \over \alpha (2 + \alpha)
(28 + 8 \alpha + \alpha^2)} \phi_{(0)}^3,\\
&=& \phi_{(0)}\Big(1 - {3 (6 + \alpha)\over \alpha (2 + \alpha)
(28 + 8 \alpha + \alpha^2)}C(\alpha)^2 a^{6/(2+\alpha)}\Big),\nonumber
\label{eq:phitildea}  
\end{eqnarray}
\begin{eqnarray}
\widetilde{\rho}_\phi &=&
\rho_{\phi(0)} + \rho_{\phi(1)},\nonumber\\
\label{eq:rhotildea}
&=& \rho_{\phi (0)}\Big(1 -\frac{\alpha(4+\alpha)} {6+\alpha}\Big({\phi_{(1)}\over 
\phi_{(0)}}\Big)\Big),\\
&=& \rho_{\phi (0)}\Big(1 + { 3(4+\alpha)\over (2+\alpha)(28 + 8 \alpha + \alpha^2)}
C(\alpha)^2 a^{6/(2+\alpha)}\Big),\nonumber
\end{eqnarray}
and
\begin{eqnarray}
\widetilde w &=&
w_{(0)} + w_{(1)},\nonumber \\
\label{eq:wphitildea}
 &=& w_{(0)}\Big(1 - \frac{\alpha(4+\alpha)} {6+\alpha}\Big({\phi_{(1)}\over 
\phi_{(0)}}\Big)\Big),\\
&=& w_{(0)}\Big(1 +  { 3(4+\alpha)\over (2+\alpha)(28 + 8 \alpha + \alpha^2)}
C(\alpha)^2a^{6/(2+\alpha)}\Big).\nonumber
\end{eqnarray}
We also define
$\widetilde{\Omega}_{\phi} = \widetilde{\rho}_{\phi}/(\widetilde{\rho}_{\phi} + \rho_{m})$.

\section{\label{sec:NUM}A comparison of Analytic and Numerical Results}

To test the accuracy of our analytical predictions, 
we numerically integrate equation~(\ref{eq:eom})
for the cases $\alpha = 1$, $\alpha = 2$, and $\alpha = 4$.
In order to compare analytic and numerical results for a specific case, we adjust
$\kappa$ to give
$\Omega_{m0} = 0.3$ and $\Omega_{\phi0} = 0.7$.
In figures $1 - 4$ below, the zeroth-order (tracker) and 
first-order analytic solutions for $\phi$, $\rho_{\phi}$, $\Omega_{\phi}$, 
and $w$ are compared to their numerical counterparts.

In Fig. 1, we present
the ratios of the tracker and first-order quintessence fields
to the true (numerical) field: $\phi_{(0)}/\phi$,
and $\widetilde{\phi}/\phi$, respectively.
As expected, $\widetilde{\phi}$ agrees exceptionally well with the true evolution at early
times, when the quintessence density is sub-dominant;
the error for all three cases is less than 2\% for $z \ge 1$.
The analytic expression begins to break down at later times, becoming
significantly less accurate at
$z = 0$, but it is still much more representative of the true behavior
than the tracker solution $\phi_{(0)}$.
\bfig
\includegraphics[height = .35\textheight, width = .5\textwidth]{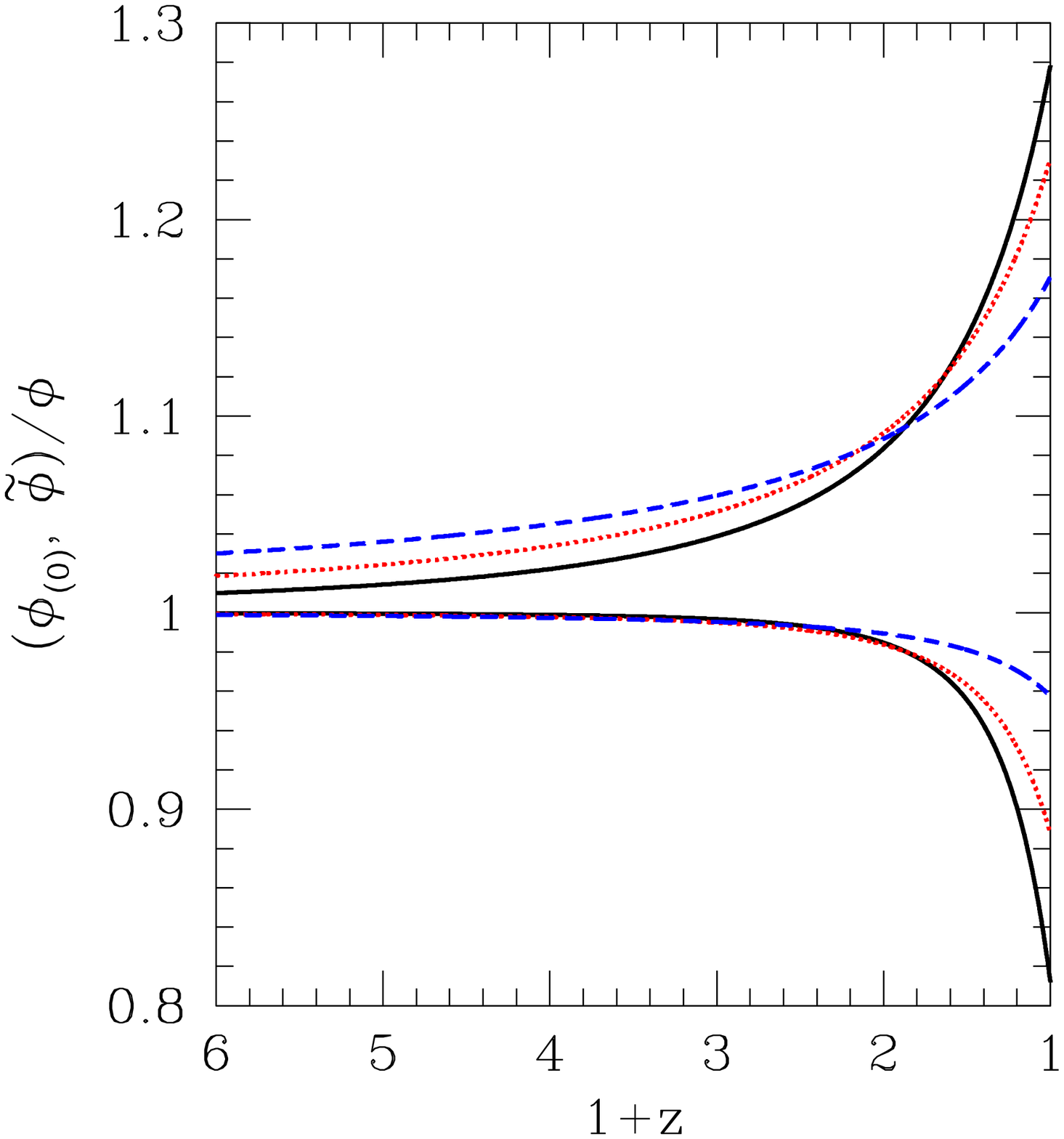}
\caption{
The ratios $\phi_{(0)}/\phi$ (upper three curves) and 
$\widetilde \phi/\phi$ (lower three curves)
for $\alpha = 1$ (black, solid), $\alpha = 2$ (red, dotted), 
and $\alpha = 4$ (blue, dashed).}
\efig

In Fig. 2, we present
$\rho_{\phi(0)}/\rho_\phi$,
and $\widetilde\rho_{\phi}/\rho_\phi$.
\bfig
\includegraphics[height = .35\textheight, width = .5\textwidth]{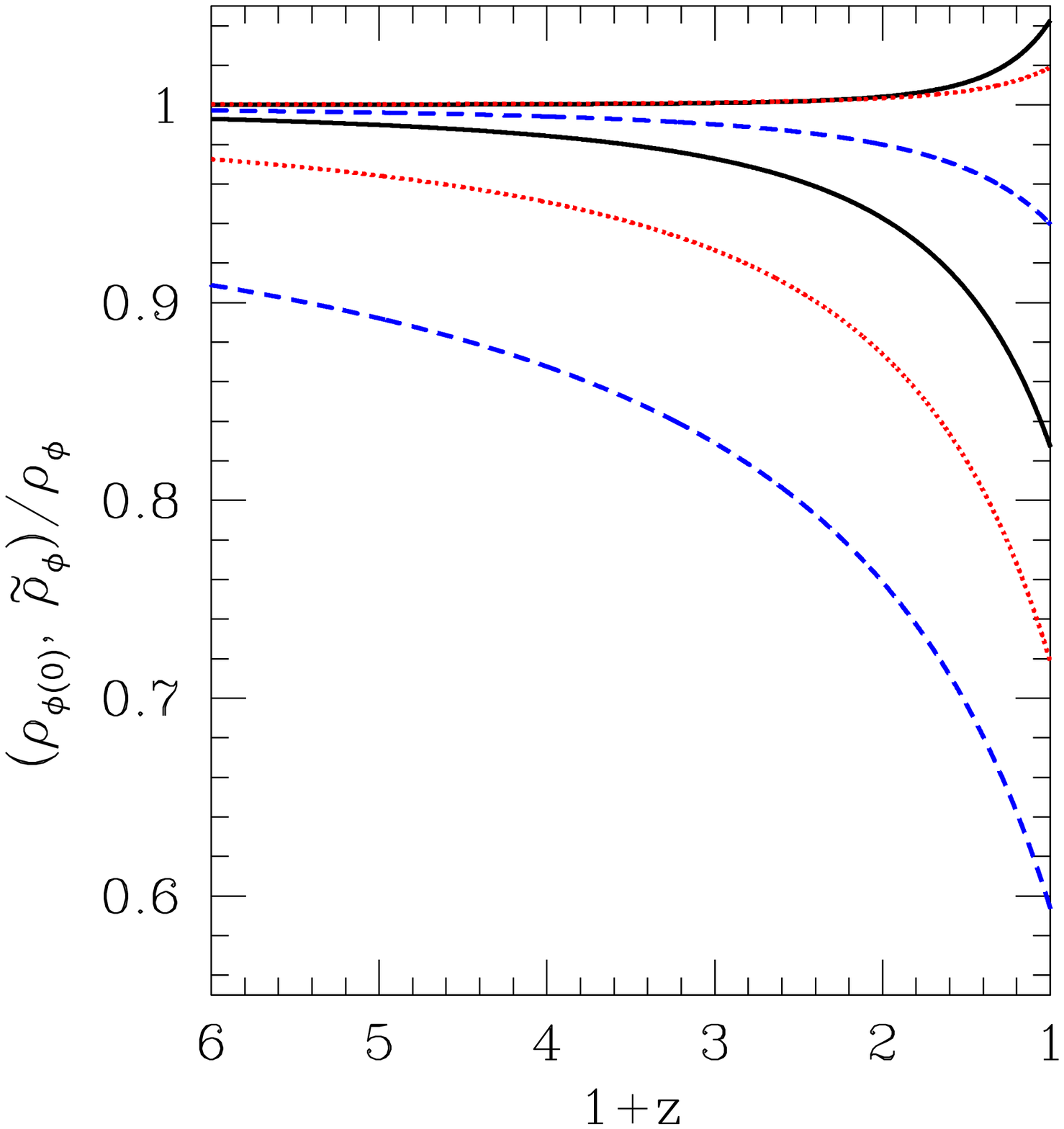}
\caption{
The ratios $\rho_{\phi(0)}/\rho_\phi$ (lower three curves) and 
$\widetilde \rho_{\phi}/\rho_\phi$ (upper three curves) 
for $\alpha = 1$ (black, solid), $\alpha = 2$ (red, dotted),
and $\alpha = 4$ (blue, dashed).}
\efig
Here the power of the first-order approach is evident.  The first-order solutions in these
cases are accurate to within 6\% all of the way down to $z = 0$.  In contrast,
the tracker expression for $\rho_\phi$ grossly underestimates
the quintessence density at low redshift.

In Fig. 3, we give the ratios $\Omega_{\phi (0)}/\Omega_{\phi}$ and 
$\widetilde{\Omega}_{\phi}/\Omega_{\phi}$.
\bfig
\includegraphics[height = .35\textheight, width = .5\textwidth]{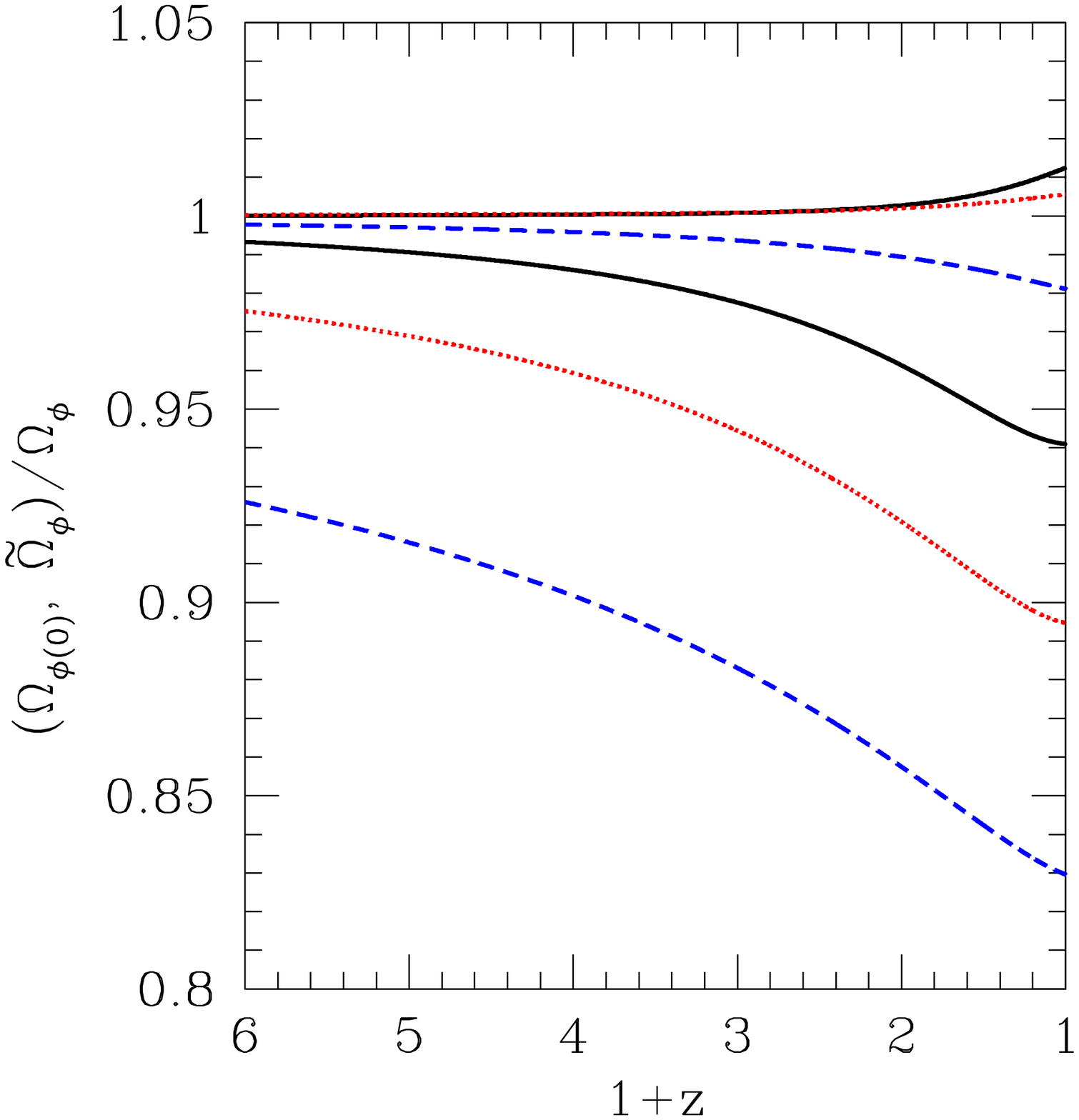}
\caption{
The ratios $\Omega_{\phi(0)}/\Omega_\phi$ (lower three curves) and 
$\widetilde\Omega_{\phi}/\Omega_\phi$ (upper three curves) 
for $\alpha = 1$ (black, solid), $\alpha = 2$ (red, dotted),
and $\alpha = 4$ (blue, dashed).}
\efig
The agreement between $\widetilde{\Omega}_\phi$ and
the true $\Omega_\phi$ is quite remarkable.  Down to $z=1$,
the error in the first-order expression is less than 1\% for all
three cases we have examined, and even at $z = 0$,
the error is less than 2\%.

Finally, we consider the equation of state parameter, $w$.  In Fig. 4,
we show the ratio of the tracker value $w_{(0)}$ (which is, of course,
constant) to the true $w$ and $\widetilde{w}/w$. 
\bfigt
\includegraphics[height = .35\textheight, width = .5\textwidth]{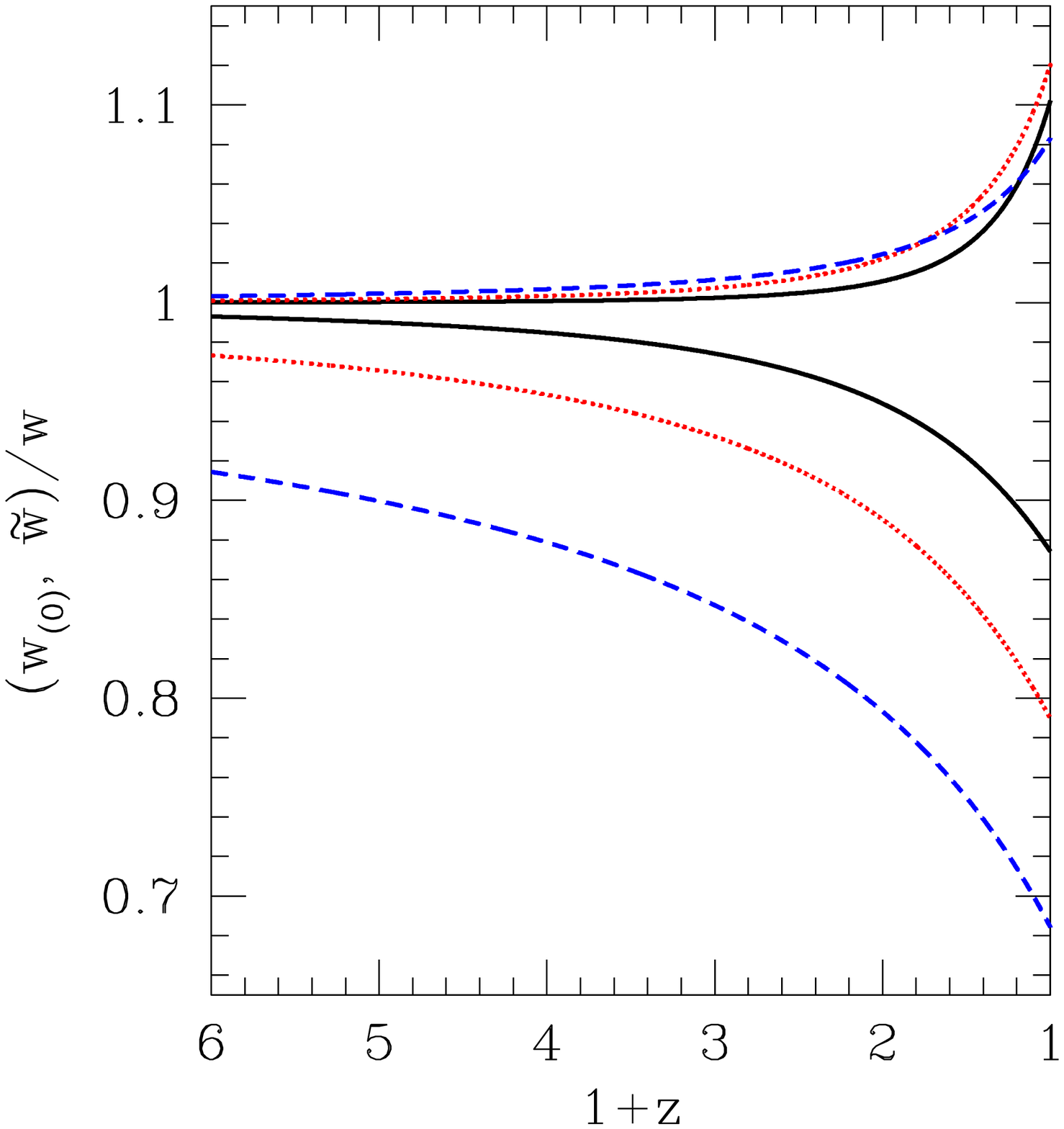}
\caption{
The ratios $w_{(0)}/w$ (lower three curves) and 
$\widetilde w/w$ (upper three curves) 
for $\alpha = 1$ (black, solid), $\alpha = 2$ (red, dotted),
and $\alpha = 4$ (blue, dashed).}
\efig
Agreement between $\widetilde{w}$ and $w$ is quite good at early times, with an error
of less than 2\% for $z \ge 1$, increasing to 4\% at $z=0.5$ and
10\% at $z=0$.

\section{\label{sec:CON}Discussion}

Our first-order corrections to the pure matter-dominated solutions for $\phi$,
$\rho_{\phi}$, $\Omega_{\phi}$, and $w$
characterize the evolution of these quantities extremely well
at $1+z>2$.  Although, unsurprisingly, the first-order solutions begin
to break down for $1+z < 2$, when the quintessence component dominates
the expansion, they are suprisingly accurate even to the present.

We can draw some simple qualititative conclusions from these results.
It is instructive to use equations (\ref{phidensity}) and
(\ref{eq:wphitildea}) to express ${w}$ as a function of
$\Omega_\phi$, since these are the two directly-observable
quantities of interest.  We find, to
first order in $\rho_{\phi (0)}/\rho_m$:
\begin{equation}
\label{wOmega1}
{w} = 
-{2 \over 2+ \alpha} - {2 \alpha (4 + \alpha) \over (2+\alpha)(28 + 8 \alpha +
\alpha^2)} \Omega_\phi.
\end{equation}
The first term in this equation is just $w_{(0)}$,
while the second term
shows how
${w}$ diverges from the tracker value at early times,
when the quintessence field is just starting to contribute
to the expansion of the universe.  We see that the deviation of ${w}$ from
$w_{(0)}$ depends linearly on $\Omega_\phi$.  Further, the constant
of proportionality is a slowly-varying function of $\alpha$, ranging from
0.09 for $\alpha = 1$ to 0.14 for $\alpha = 4$.  Hence, the rate at which
$w$ deviates from $w_{(0)}$ at early times is almost independent
of $\alpha$.  
Equation (\ref{wOmega1}) is an excellent fit for small $\Omega_\phi$, but
it works suprisingly well even up to the present.  For $\alpha = 1-4$,
the error in the prediction for $w$ is less than 10\%, with smaller
errors for smaller (and more physically relevant) values of $\alpha$.

A number of approximations have been proposed to simulate the evolution
of $w$ at late times.  For arbitrary
quintessence models, a common approximation is a \linebreak[4]
Taylor expansion in $z$ \cite{astier,goliath,maor,huterer,weller}:
\begin{equation}
\label{wlin}
w_{\rm lin} = w_0 + w_1 z + ...,
\end{equation}
while more complex parametrizations have been examined by
Bassett et al.\cite{bassett} and by Corasaniti and
Copeland \cite{corasaniti}.

For the specific
case of inverse power-law models, Efstathiou \cite{GE} showed that
a good empirical fit to the time-evolution of $w$ is:
\begin{equation}
\label{wE}
w_{\rm Efst} = w_0 + w_1 \ln (1+z).
\end{equation}
Our results suggest an alternative parametrization.  We can write equation
(\ref{eq:wphitildea}) in the form
\begin{equation}
\label{wfit}
w_{\rm fit} = w_{(0)} + (w_0 - w_{(0)})(1+z)^{-6/(2+\alpha)},
\end{equation}
Since $w_{(0)}$ is given by
equation (\ref{w(0)}),
equation (\ref{wfit}) has only one undetermined parameter, the present value
of the equation of state, $w_0$, which can be determined numerically and used 
to derive a fit to $w$. As shown in Fig. 5, the error introduced by this 
\bfig
\includegraphics[height = .35\textheight, width = .5\textwidth]{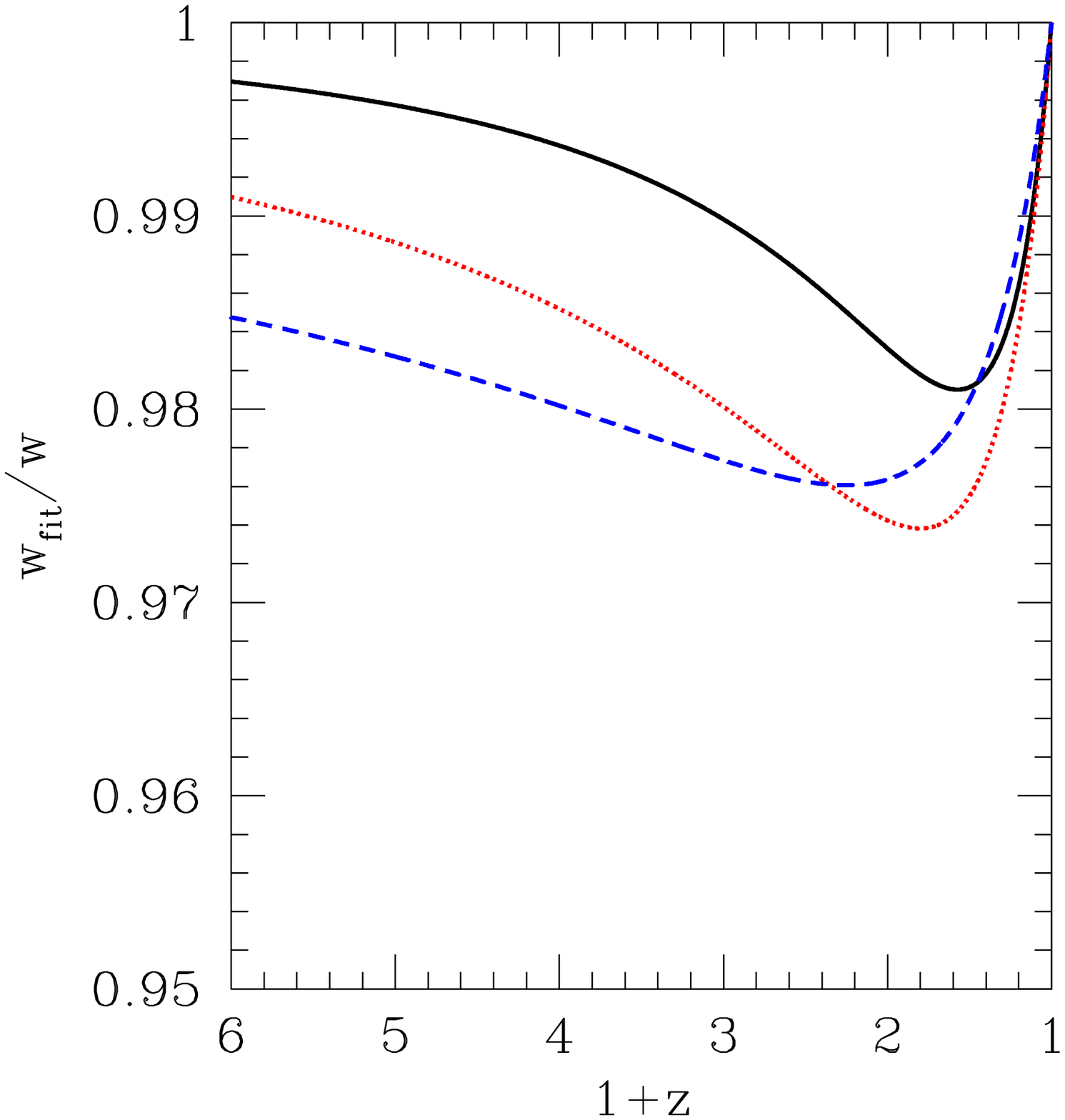}
\caption{
The ratio $w_{\rm fit}/w$, where $w_{\rm fit}$ is given by equation
(\ref{wfit}), 
for $\alpha = 1$ (black, solid), $\alpha = 2$ (red, dotted),
and $\alpha = 4$ (blue, dashed).}
\efig
\bfigt
\includegraphics[height = .35\textheight, width = .5\textwidth]{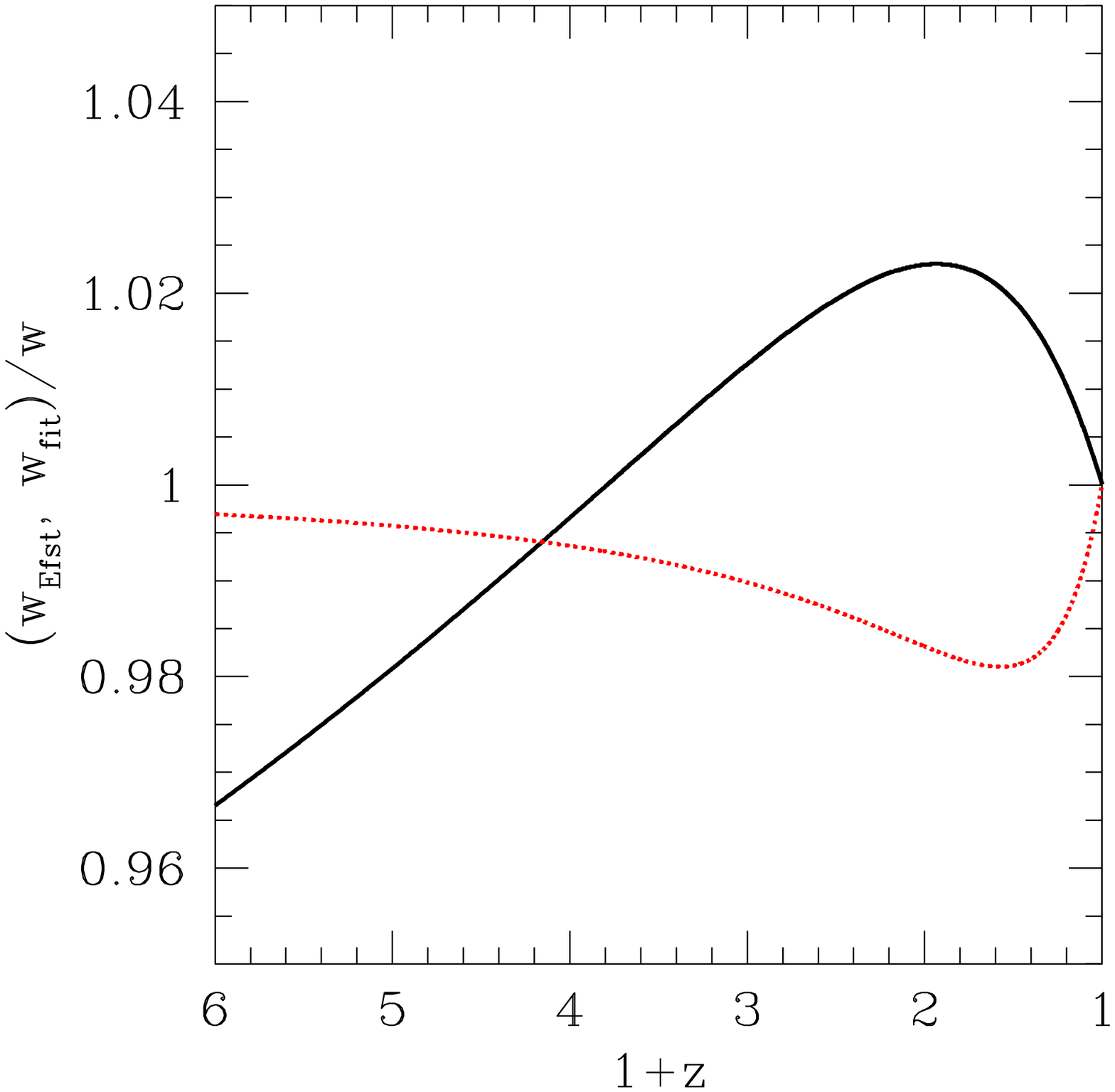}
\caption{
The ratios $w_{\rm Efst}/w$ (black, solid), and $w_{\rm fit}/w$ (red, dotted) for the $\alpha = 1$ case.} 
\efig
approximation is less than 3\% for $\alpha$ in the range $1-4$.

To show how our fit (equation \ref{wfit}) compares to the empirical fit of equation (\ref{wE}), we have plotted their behavior for
the $\alpha = 1$ case in
Fig. 6. A $\chi^2$ minimization routine was used to determine the best-fit value of $w_1$ over the
plotted redshift range for $w_{\rm Efst}$.  (In both approximations, we set
$w_0$ equal to the present-day value of the equation of state that is found numerically for $\alpha=1$). Despite the
fact that $w_{\rm Efst}$ contains one more fitted parameter than $w_{\rm fit}$, it is clear from Fig. 6
that $w_{\rm fit}$ agrees better with the numerical
behavior of $w$ at high redshift and shows comparable accuracy at low redshift.  Note,
however, that if we restrict the redshift range to much lower $z$
(e.g., $0 < z <1$), and recalculate $w_1$ to give the
best fit for $w_{\rm Efst}$ only within this redshift range, then $w_{\rm Efst}$
gives better agreement than $w_{\rm fit}$, although both expressions are accurate
to within 2\% over this redshift interval.

Neither of these results is unexpected.  The expression for $w_{\rm fit}$ in equation (\ref{wfit})
is derived by perturbing the
high-redshift evolution of $\phi$, so we expect it to be more accurate at higher redshifts than $w_{\rm Efst}$.  On the other
hand, the expression for $w_{\rm Efst}$ was derived from an empirical fit at low $z$, and so it is no surprise that this expression
can be made to work better at low redshifts.

While we have tested our results against a fiducial flat model with
$\Omega_{m 0} = 0.3$, nothing in our analysis depends on this
assumption.  For obvious reasons, the agreement between our
analytic results and the true evolution of $\rho_\phi$ and $w$
will be better than the ones presented here if $\Omega_{m 0} > 0.3$,
and worse if $\Omega_{m 0} < 0.3$.

Although we have concentrated specifically on the inverse power-law potentials
in this paper, it is obvious that the techniques exploited here can be
applied to any quintessence model in which the quintessence energy
density is subdominant at early times and dominates at low redshift.

\section{Acknowledgments}

R.J.S. was supported in part by the Department
of Energy (DE-FG02-91ER40690).

\end{document}